\documentclass[twocolumn,preprint]{aastex63}
\usepackage{amsmath}
\usepackage{color}
\hypersetup{linkcolor=red,citecolor=blue}

\shortauthors{Huang et al.}

\usepackage{graphicx}
\graphicspath{{fig/}}
\usepackage{amsmath}
\usepackage{amssymb}
\usepackage{verbatim}
\usepackage{enumerate}

\begin{document} 
\title{Two-Stage Nature of a Solar Flare with Parallel and Semi-Circular Ribbons}
\correspondingauthor{Hao Ning; Ze Zhong}\email{haoning@sdu.edu.cn; zezhong@sdu.edu.cn}

\author[0009-0004-1653-1735]{Ruifei Huang}
\affil{Center for Integrated Research on Space Science, Astronomy, and Physics, Institute of Frontier and Interdisciplinary Science, Shandong University, Qingdao 266237, China\\}
\affil{Shandong Key Laboratory of Space Environment and Exploration Technology, Institute of Space Sciences, Shandong University, Weihai 264209, China\\}

\author[0000-0001-8132-5357]{Hao Ning}
\affil{Center for Integrated Research on Space Science, Astronomy, and Physics, Institute of Frontier and Interdisciplinary Science, Shandong University, Qingdao 266237, China\\}
\affil{Shandong Key Laboratory of Space Environment and Exploration Technology, Institute of Space Sciences, Shandong University, Weihai 264209, China\\}

\author[0000-0001-5483-6047]{Ze Zhong}
\affil{Center for Integrated Research on Space Science, Astronomy, and Physics, Institute of Frontier and Interdisciplinary Science, Shandong University, Qingdao 266237, China\\}
\affil{Shandong Key Laboratory of Space Environment and Exploration Technology, Institute of Space Sciences, Shandong University, Weihai 264209, China\\}
\affil{School of Astronomy and Space Science and Key Laboratory of Modern Astronomy and Astrophysics, Nanjing University, Nanjing 210023, China\\}

\author[0000-0002-1190-0173]{Ye Qiu}
\affil{Institute of Science and Technology for Deep Space Exploration, Suzhou Campus, Nanjing University, Suzhou 215163, China\\}

\author[0000-0003-4804-5673]{Zhenyong Hou}
\affil{School of Earth and Space Sciences, Peking University, Beijing 100871, China\\}

\author[0000-0002-4241-9921]{Yang Su}
\affil{Key Laboratory of Dark Matter and Space Astronomy, Purple Mountain Observatory, CAS, Nanjing 210023, China\\}

\author[0000-0001-7693-4908]{Chuan Li}
\affil{School of Astronomy and Space Science and Key Laboratory of Modern Astronomy and Astrophysics, Nanjing University, Nanjing 210023, China\\}
\affil{Institute of Science and Technology for Deep Space Exploration, Suzhou Campus, Nanjing University, Suzhou 215163, China\\}

\author[0000-0003-1034-5857]{Xiangliang Kong}
\affil{Shandong Key Laboratory of Space Environment and Exploration Technology, Institute of Space Sciences, Shandong University, Weihai 264209, China\\}
\affil{Center for Integrated Research on Space Science, Astronomy, and Physics, Institute of Frontier and Interdisciplinary Science, Shandong University, Qingdao 266237, China\\}

\author[0000-0001-6449-8838]{Yao Chen}
\affil{Center for Integrated Research on Space Science, Astronomy, and Physics, Institute of Frontier and Interdisciplinary Science, Shandong University, Qingdao 266237, China\\}
\affil{Shandong Key Laboratory of Space Environment and Exploration Technology, Institute of Space Sciences, Shandong University, Weihai 264209, China\\}

%------**********************************--------------
\begin{abstract}
	
Flare ribbons with parallel and circular morphologies are typically associated with different magnetic reconnection models, and the simultaneous observation of both types in a single event remains rare. Using multi-wavelength observations from a tandem of instruments, we present an M8.2-class flare that occurred on 2023 September 20, which produced quasi-parallel and semi-circular ribbons. The complex evolution of the flare includes two distinct brightening episodes in the quasi-parallel ribbons, corresponding to the two major peaks in the hard X-ray (HXR) light curve. In contrast, the brightening of semi-circular ribbons temporally coincides with the local minimum between the two peaks. Using potential field extrapolation, we reconstruct an incomplete dome-like magnetic structure with a negative polarity embedded within the northwestern part of the semi-circular positive polarity. Consequently, the magnetic configuration comprises two sets of field lines with distinct magnetic connectivities.	
We suggest that the standard flare reconnection accounts for the two-stage brightening of quasi-parallel ribbons associated with the two HXR peaks. 
Between the two stages, this process is constrained by the interaction of eruptive structures with the dome. The interaction drives the quasi-separatrix layer reconnection, leading to the brightening of semi-circular ribbons. It also suppresses the standard flare reconnection, resulting in a delayed second HXR peak.

\end{abstract}


%------**********************************--------------
\section{Introduction} \label{Sect.1}

Solar flare represents one of the most eruptive processes in solar atmosphere with abrupt enhancements of radiation in a wide range of wavelengths \citep[see, e.g.,][and references therein]{2008Benz}. It's widely acknowledged that the magnetic reconnection plays an important role in energy release and particle acceleration \citep{Parker1957,Sweet1958,PriestForbes2002,2011Shibata, Su2013}. Typical two ribbon flares can be well explained by the 2.5 dimensional standard flare model known as the CSHKP model \citep{1964Carmichael,1968Sturrock,1974Hirayama,1976Kopp}. As the flux rope rises and stretches the overlying magnetic field lines, current sheets form between oppositely directed fields, where magnetic reconnection occurs and electrons are energized. These energetic electrons flow downwards, producing bright flare ribbons at the footpoints. As reconnection progresses, outer magnetic field lines get involved, leading to the observable separation of flare ribbons. 

There is a special type of flares known as circular ribbon flares, associated with three-dimensional (3D) null-point fan-spine magnetic topology \citep[e.g.,][]{2009Masson,2012reid,2012Wang,Sun2013,2016Zhang,2017Masson,Li2018,2019Zhong,2020Lee,2020Liu,2022Zhang,2022Ning,2023Mitra}. The null point divides the spine into two parts with distinct magnetic connectivities \citep{1996Parnell}, i.e., the inner and outer spines linking the main flaring region and the remote area, respectively \citep{2009Masson}. The energetic electrons accelerated via the null point reconnection would also penetrate downward along the magnetic field lines on the fan surface, heating the dense plasmas in chromosphere and leading to the brightening of circular-shaped ribbons. 

Magnetic reconnection also occurs within the quasi-separatrix layers  \citep[QSLs;][]{1995Demoulin,1996Demoulin}. These regions exhibit strong gradients in the magnetic field connectivity, quantified by the squashing factor Q \citep{2002Titov}. In particular, the reconnection occurs preferentially at locations where Q reaches local maximum. The theory of QSL reconnection has been applied to interpret various observational features of solar flares \citep[e.g.,][]{2015Li,2017Jing,2022Pan}. This provides insights into the 3D dynamics of magnetic reconnection.

During solar flares, hard X-ray (HXR) emissions are generally produced by bremsstrahlung of energetic electrons penetrating into the footpoints, co-spatial with the flare ribbons, followed by the chromospheric evaporation \citep{Li2015, Tian2015}.  HXR emissions are closely related to electron acceleration during solar flares, serving as a crucial tool for studying the energy release process in magnetic reconnection. A comprehensive multi-wavelength observational analysis is required to study the dynamic evolution of flares in different aspects.

According to previous observational studies \citep[e.g.,][]{2012reid,2012Wang,Sun2013,2017Masson,Li2018,2019Zhong,2020Lee,2023Mitra}, a majority of circular ribbon flares are confined events, with only a few accompanied by coronal mass ejections (CMEs) \citep{2018Song,2019Liu,2020Liu,Joshi2021,2021Ibrahim}. 
Under certain conditions, both circular ribbons and parallel ribbons are observed in the same event \citep[e.g.,][]{2015yangkai,2017Hernandez}, with the rising flux rope triggering reconnection in different topologies. The brightening sequence of different sets of ribbons varies from event to event. For example, \cite{Joshi2015} studied a flare where the parallel ribbons formed before the circular ribbons, while \cite{Joshi2021} documented a reverse case. \cite{2023Mitra} reported an event with the onset of circular and parallel ribbons taking place almost simultaneously. It is not clear how the reconnections in different topologies affect the time evolution of ribbons exhibiting different morphologies. More observations are required to verify and refine the 3D models of flares with complicated ribbons.

In this paper, we present a multi-wavelength analysis of a complicated flare exhibiting quasi-parallel and semi-circular ribbons. This flare features two types of reconnection processes: (1) the reconnection known from the standard CSHKP flare model in 2D, which produces a pair of ribbons and is referred to as standard flare reconnection for simplicity, and (2) QSL reconnection in the fan-spine topology. These produce the temporal evolution of HXR emission with two peaks and a local minimum. The observational data and event overview are introduced in Section \ref{Sect.2}. The analysis of multi-wavelength data is presented in Section \ref{Sect.3}. Magnetic configuration and a two-stage scenario are described in Section \ref{Sect.4}, followed by a summary and discussion in Section \ref{Sect.5}.
%------**********************************--------------

\section{Observational data and event overview}\label{Sect.2}

\begin{figure}[htbp]
	\centering
	\includegraphics[width=0.7\textwidth]{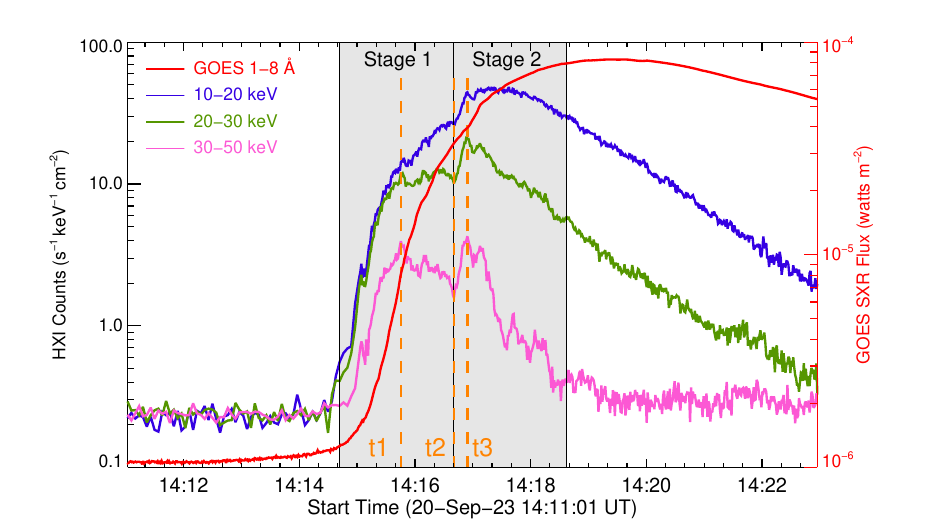}
	\caption{(a) GOES soft X-ray fluxes at 1--8 {\AA} (red) and ASO-S/HXI X-ray light curves in 10--20 (blue), 20--30 (olive), and 30--50 (pink) keV. Three vertical dashed lines represent the moments of t1 (14:15:44 UT), t2 (14:16:40 UT), and t3 (14:16:53 UT), corresponding to the two major peaks and the trough of the HXR light curves in 30--50 keV. The two stages of this event (stage 1: 14:14:41--14:16:40 UT; stage 2: 14:16:40--14:18:36 UT) are shaded in gray, separated by t2.}
	\label{figure01}
\end{figure}

On 2023 September 20, an eruptive flare occurred in the active region 13435 (N07, E35). According to the soft X-ray instrument onboard the Geostationary Operational Environmental Satellite (GOES), the M8.2-class flare started at 14:11 UT, peaked at 14:19 UT, and ended at 14:25 UT (Figure~\ref{figure01}). The flare was associated with an extreme ultraviolet (EUV) wave and a Moreton wave \citep{2025Zhong}, followed by a coronal mass ejection (CME), which was recorded by the Large Angle and Spectrometric Coronagraph (LASCO) C2 onboard the Solar Heliospheric Observatory \citep[SOHO;][]{1995Brueckner}. 

To study this event, we utilize observational data collected by a tandem of space-based instruments, including the Atmosphere Imaging Assembly \citep[AIA;][]{2012Lemen} onboard the Solar Dynamics Observatory \citep[\textit{SDO};][]{2012Pesnell}, the Helioseismic and Magnetic Imager \citep[HMI;][]{2012Scherrer} onboard \textit{SDO}, the Hard X-ray Imager \citep[HXI;][]{2019Zhang,2019Su} onboard the Advanced Space-based Solar Observatory \citep[ASO-S;][]{2019Gan}, the Solar Upper Transition Region Imager \citep[SUTRI;][]{2023Bai,2023Wang} onboard the Space Advanced Technology demonstration satellite (SATech-01), together with the H$\alpha$ imaging spectrograph \citep[HIS;][]{2022Liu} onboard the Chinese H$\alpha$ Solar Explorer \citep[CHASE;][]{2019Li,2022Li}.

AIA provides high-cadence EUV (12 s) and UV (24 s) data, across a wide temperature range. The spatial resolution is $\sim 0.6\arcsec$ per pixel.
HXI is designed to observe HXR emission ranging from 15 to 300 keV with a spatial resolution of $3\arcsec$ per pixel and an unprecedented high cadence (as high as 0.125 s). SUTRI provides Ne \uppercase\expandafter{\romannumeral7} 46.5 nm spectral line emitted from the transition region \citep{2017Tian} with a spatial resolution of $\sim 1.2\arcsec$ per pixel and a temporal cadence of 30 s. CHASE firstly obtains full-disk spectroscopic H$\alpha$ (6559.7--6565.9 {\AA}) and Fe \uppercase\expandafter{\romannumeral1} (6567.8--6570.6 {\AA}) emission and derived dopplergrams with a temporal resolution of 60 s and a spatial resolution of $1.04\arcsec$ per pixel of the binning mode after calibration \citep{2022Qiu}.

Figure~\ref{figure01} presents the HXR light curves observed by ASO-S/HXI. The time profile of 10--20 keV energy band exhibiting gradual rise and decay phases is similar to that of the GOES soft X-ray (SXR) flux. The fluxes in higher energy bands (20--30 and 30--50 keV) display multiple peaks. The 30--50 keV light curve shows two major peaks at 14:15:44 (t1) and 14:16:53 UT (t3), separated by a local minimum at 14:16:40 UT (t2). We categorize the temporal evolution into two stages: stage 1 (14:14:41--14:16:40 UT), and stage 2 (14:16:40--14:18:36 UT).

In stage 1, the fluxes in the whole energy bands show an impulsive rise. The fluxes in 20--30 and 30--50 keV reach the first peak at t1, followed by fluctuations with a minor peak at $\sim$14:16:10 UT. Subsequently, the 30--50 keV flux decreases to a trough (local minimum) at t2.
In stage 2, the 20--30 and 30--50 keV fluxes undergo a second sharp rise, reaching the second major peak at t3, and then gradually decrease. A minor peak occurs at $\sim$14:17:10 UT. At the end of this stage, the 30--50 keV flux decays to the pre-flare background level. The multi-peak HXR profile indicates complex energy release processes.

Figure~\ref{figure02} displays multi-wavelength images during the impulsive phase, presenting the brightening features across different atmospheric layers.  Complicated ribbon structures observed in H$\alpha$ and 1600~\AA{} images comprise the following parts: two quasi-parallel ribbons (R1 and R2), semi-circular ribbons (R3 and R4), side ribbon (R5), and remote ribbon (R6). All light curves observed by AIA, SUTRI and GOES show a rapid rise from $\sim$14:15 UT and decay to pre-flare levels at $\sim$14:40 UT. In particular, the peaks of the 304 and 465 {\AA} light curves coincide with the second HXR peak at t3. The 131 {\AA} emission peaks first at $\sim$14:20:10 UT, and the 94 {\AA} at $\sim$14:20:38 UT. 
Note that two distinct peaks occur in the 1600 {\AA} at 14:15:50 UT and 14:17:02 UT, temporally consistent with the two HXR peaks.

%------**********************************--------------
\section{Analysis of multi-wavelength data}\label{Sect.3}
%------**********************************--------------
%ribbon
\subsection{Analysis of (E)UV data}

As shown in Section~\ref{Sect.2}, the flare ribbons exhibit four morphological groups. This section focuses on the  spatio-temporal evolution of the different flare ribbons and associated loop systems using (E)UV data.

\subsubsection{Analysis of flare ribbons and underlying magnetic field}
\begin{figure}[htbp]
	\centering
	\includegraphics[width=0.95\textwidth]{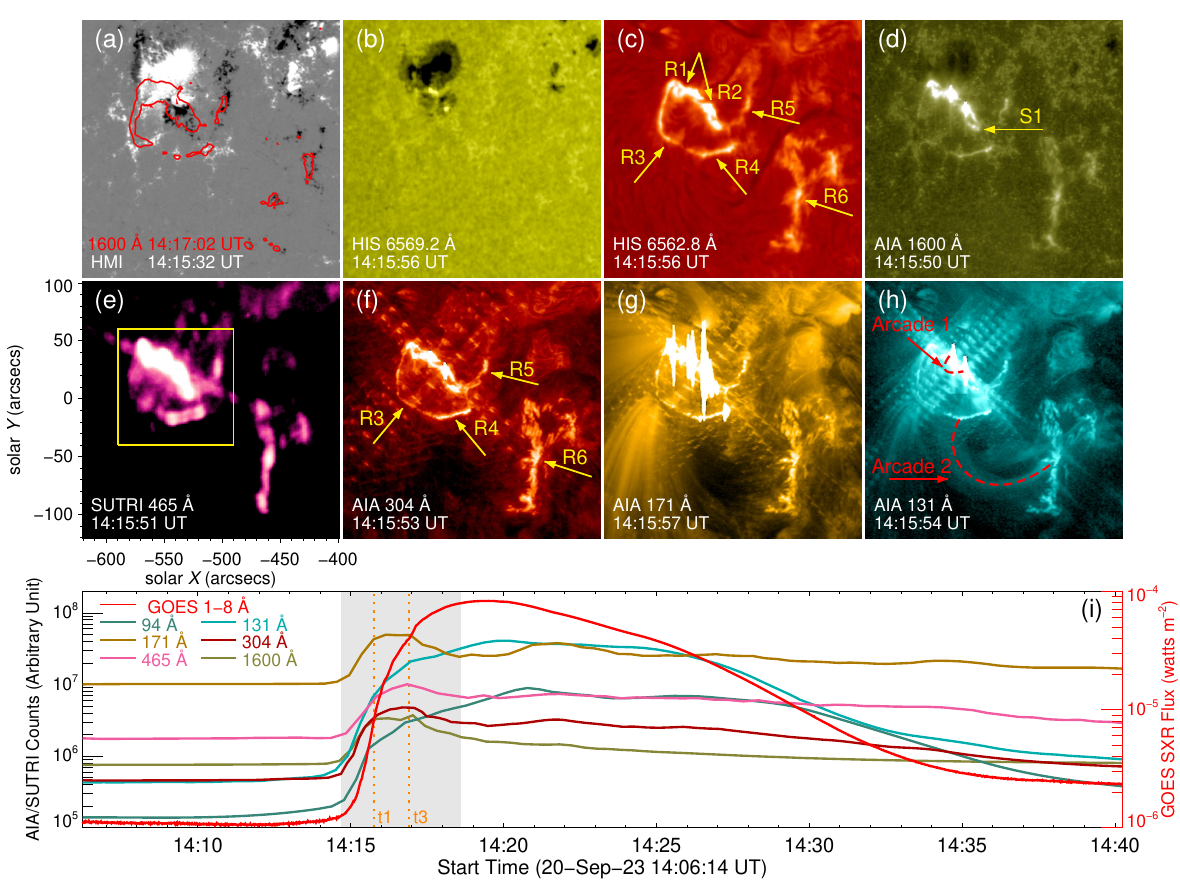}
	\caption{An overview of the M8.2-class flare on 2023 September 20. (a) \textit{SDO}/HMI line-of-sight magnetogram of the flaring region, overlaid with red contours at the 90\% intensity level from the \textit{SDO}/AIA 1600~\AA{} emission. (b)--(h) Images observed at $\sim$14:15:44 UT (t1) showing CHASE Fe I (6569.2 {\AA}), H$\alpha$ (6562.8 {\AA}), AIA 1600 {\AA}, SUTRI 465 {\AA}, AIA 304 {\AA}, 171 {\AA}, and 131 {\AA} emissions. R1--R6 refer to the flare ribbons. The yellow arrow in panel (d) points to ribbon S1. Arcades 1 and 2 are marked by red dashed lines in panel (h). (i) Light curves of GOES SXR flux at 1--8 {\AA} (red), SUTRI 465 {\AA} (pink), and AIA 94 {\AA} (atrovirens), 131 {\AA} (blue), 171 {\AA} (brown), 304 {\AA} (deep red), and 1600 {\AA} (green), respectively. The AIA and SUTRI fluxes are integrated over the yellow rectangle indicated in panel (e). The light gray shading denotes the same time interval as that in Figure~\ref{figure01}. An animation of this figure is available.}
	\label{figure02}
\end{figure}

\begin{figure}[htbp]
	\centering
	\includegraphics[width=0.95\textwidth]{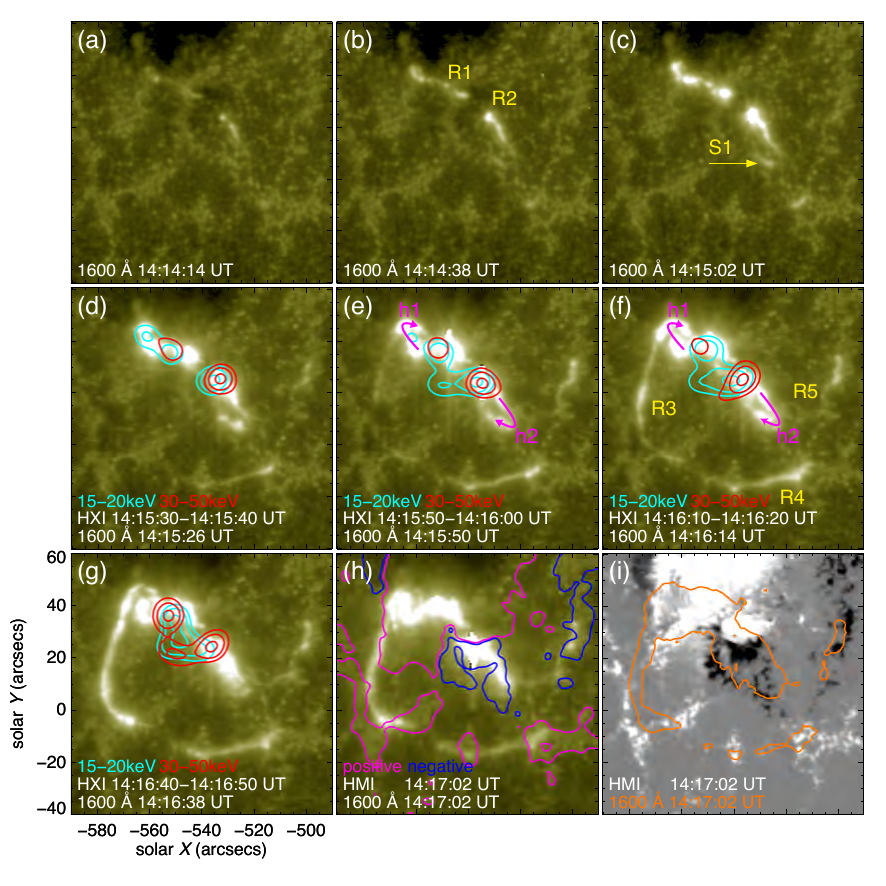}
	\caption{(a)--(h) AIA 1600 {\AA} images from 14:14:14 to 14:17:02 UT. Cyan and red contours in panels (d)--(g) show HXR sources in 15--20 and 30--50 keV, respectively, reconstructed via the CLEAN algorithm and plotted at 40\%, 60\%, and 90\% of the maximum flux. In panel (h), the image is overlaid by the pink and blue contours of the HMI line-of-sight magnetogram being $\pm$ 80 G. Arrows h1 and h2 in panels (e) and (f) mark the hook-shaped extension of ribbons R1 and R2. (i) HMI line-of-sight magnetogram at 14:17:02~UT. Orange contours represent 90\% of the maximum intensity from the co-temporal AIA 1600~\AA{} image. An animation of this figure is available. }
	\label{figure03}
\end{figure}

Figure~\ref{figure03} and the accompanying animation display the detailed evolution of the flare ribbons at 1600 {\AA}. The ribbons R1 and R2 began brightening at 14:14:38~UT. After 14:15:26 UT, the inner ribbon (S1), the remote ribbon (R6), as well as the northern ends of the side ribbon (R5) and the semi-circular ribbons (R3 and R4) brightened faintly (see the accompanying animation in Figure~\ref{figure02}). Then, around 14:15:50 UT, the northern end of ribbon R1 and the southern end of ribbon R2 connected with ribbons R3 and S1, respectively, forming two oppositely oriented hooked structures, as outlined by h1 and h2. Finally, between 14:16:14 and 14:17:02 UT, the brightness of ribbon R3 increased significantly, propagating counter-clockwise, while ribbon R4 propagated in the opposite direction.

During this flare, ribbons R1 and R2 went through a two-stage brightening, which saturated in the 1600 {\AA} images at 14:15:50 and 14:17:02 UT, respectively. However, the other ribbons reached their peak intensity between these two moments. This temporal evolution is also visible in the H$\alpha$, 465 {\AA}, and 304 {\AA} images. Furthermore, the ribbons R1 and R2 left their imprint in the photosphere at 14:15:56 UT (see Figure~\ref{figure02}(b)).

Figures~\ref{figure02}(a) and~\ref{figure03}(i) display the photospheric line-of-sight magnetogram overlaid with flare ribbons. The flaring core region contains two groups of intersecting polarities:

\noindent 1) Two opposite polarities, corresponding to ribbons R1 (positive polarity) and R2 (negative polarity). The ribbons are quasi-parallel to the polarity inversion line.
 
\noindent 2) The quasi-circular positive polarity enclosing a parasitic negative polarity. Ribbons R3 and R4 are co-spatial with the positive polarities, forming a semi-circular shape. Ribbons S1 and R5 correspond to the negative polarities. Ribbon R6 spatially coincides with the negative polarity in the southwest.

\subsubsection{Analysis of light curves}
Combining the HXR light curves, we analyze the 1600~{\AA} light curves in six flare ribbon regions to investigate the two-stage energy release process (Figure~\ref{figure04}). The fluxes of regions R1 and R2 exhibit two peaks at $\sim$t1 and t3 and a local minimum at $\sim$t2. This pattern is generally correlated with the HXR profile in 30--50 keV.

Each light curve of region R3--R6 reaches a single peak during this event. The flux of region R3 peaks at 14:16:38 UT, coinciding exactly with the HXR trough at t2. The fluxes of regions R4 and R6 peak at 14:16:14 UT, while region R5 reaches its peak at 14:17:02 UT.

\subsubsection{Analysis of loop system}

As shown in the 171 and 131~\AA{} images (Figure~\ref{figure02}(g) and (h)), the flaring region contains two primary sets of loop structures: Arcade 1 (connecting ribbons R1 and R2) and Arcade 2 (connecting the ribbons R4 and R6). The evolution of these loops in 131 {\AA} is displayed in Figure~\ref{figure05} and its accompanying animation. The Arcade 1 denotes the post-flare loops which exhibited a significant brightness enhancement from $\sim$14:16~UT to 14:20~UT. We also traced the dynamics of the Arcade 2 through 131~\AA{} running-difference images (Figure~\ref{figure05}(e)--(h)). The Arcade 2 began to brighten at 14:15~UT and moved southward rapidly. After 14:17~UT, it disappeared from the field of view. The expansion of the Arcade 2 suggests an ejection process. 

\subsection{Imaging and spectral analysis of the HXR data}
We analyze the HXR imaging and spectral data from ASO-S/HXI to investigate the energy release during this event. The contours in Figure~\ref{figure06}(a) display the HXR sources of 30--50 keV at t1, t2, and t3, superimposed onto the dopplergram from CHASE at 14:16 UT. At t1, the HXR emissions primarily originate from two sources located at $\sim$($-557\arcsec$, $36\arcsec$) and ($-536\arcsec$, $25\arcsec$), respectively, co-spatial with the two flare ribbons (R1 and R2). The western source is more intense than the eastern one, showing asymmetry in the two footpoints. The background dopplergram results suggest strong downward velocity up to $\sim$40 km~s$^{-1}$ in the flare ribbons, indicating a downward plasma flow. The maxima of the sources at t2 (green pluses) and t3 (orange pluses) were located near the positions of the maxima at t1 (cyan pluses). A weak loop-top source appeared at t2, and the loop-shaped source persisted at t3. The evolution of the HXR sources is shown in Figure~\ref{figure03}(d)--(g). 

Figure~\ref{figure05}(a)--(d) display the 15--20 keV HXR sources overlaid onto the AIA 131 \AA{} images from 14:15 to 14:20~UT. At 14:15:30~UT, the sources appear at two footpoints, co-spatial with ribbons R1 and R2. Subsequently, the sources evolve to interconnect and eventually merge into a single source at 14:16:30 UT. After 14:17 UT, the source is located at the looptop, and the trajectory formed by the source motion is co-spatial with Arcade 1. In Figure~\ref{figure06}(b) and the accompanying animation, we overlay the 15--20 keV HXR sources from 14:15:20 to 14:19:50 UT onto a single H$\alpha$ image. The motion of the sources follows a loop-shaped path, highlighting the formation of the loop-top source.

As demonstrated above, the HXR radiation in the high energy band mainly originates from the ribbons R1 and R2, with no signal detected from semi-circular ribbons. The evolution of the 15--20 keV  HXR sources displays the formation of post-flare loops rooted in ribbons R1 and R2. This is consistent with chromospheric evaporation driven by nonthermal electrons in the classical flare model. 

Using the standard OSPEX software \citep{2020Tolbert}, we conduct spectral fitting analysis of HXI data at t1, t2, and t3 (Figure~\ref{figure06}(c)--(e)). The photon spectra from 30 to 70 keV can be well fitted with a power-law spectrum. At the first HXR peak (t1), the spectrum shows the hardest spectral index $\gamma=$4.27 $\pm$ 0.077 and then softens to $\gamma=$5.11 $\pm$0.11 at t2. At the second peak (t3), $\gamma$ slightly recovers to 4.90 $\pm$ 0.064. This spectral evolution implies three distinct phases: (1) a highly efficient energy release process occurs at t1; (2) the non-thermal electrons get thermalized later at t2, corresponding to the local minimum between the two peaks; (3) the spectrum get slightly harder at t3, indicating a secondary, less efficient non-thermal energy release process.

%------**********************************--------------
\section{Magnetic configuration and a two-stage scenario} \label{Sect.4}
Based on the photospheric magnetogram, we extrapolate the 3D magnetic field using potential field model \citep{1981Alissandrakis} to investigate the magnetic configuration of this flare. As shown in Figure~\ref{figure07}(a), the magnetic field exhibits a dome-like structure with a null point embedded in it, showing a fan-spine topology. Notably, there are two sets of magnetic field lines with different connectivities due to the unclosed circular positive polarity. 

One set of field lines (yellow) is shown in Figure~\ref{figure07}(a), connecting the quasi-circular positive polarity and the remote negative polarity corresponds to ribbon R6. This configuration of magnetic reconnection is shown in Figure~\ref{figure07}(b), consistent with the usual concept of circular ribbon flares \citep{2009Masson,2012reid,2012Wang}. There is a peculiar set of field lines (green) in Figure~\ref{figure07}(a), and their detailed magnetic configuration is shown in Figure~\ref{figure07}(c). Two red lines transform to cyan lines through magnetic reconnection. This configuration prevents the dome-like structure from enclosing itself. 

We calculate the squashing factor Q \footnote{\url{https://github.com/Kai-E-Yang/QSL}} to determine the QSL (Figure~\ref{figure07}(a)). We find the QSL with semi-circular structure is co-spatial with the semi-circular ribbons R3 and R4. On the northwestern side of the semi-circular ribbons, the QSL structure with negative polarity is co-spatial with the side ribbon R5.

Figure~\ref{figure08} proposes a plausible scenario to interpret the evolution of this event, involving two magnetic reconnection models: 1) the standard flare reconnection of the standard CSHKP flare model; 2) the QSL reconnection in the fan-spine topology.

In the early phase of stage 1 (Figure~\ref{figure08}(a)), the standard flare reconnection initiated the impulsive phase of the flare, resulting in the brightening of ribbons R1 and R2 with two oppositely-oriented hooked extensions. The HXR emissions originated from the footpoint sources rose simultaneously, and exhibited a hard power-law spectrum. This is consistent with the standard CSHKP flare model. Despite the absence of hot channels or sigmoids in AIA observations, we propose that a flux rope existed in this stage, and its footpoints were surrounded by the brightened hooks. The flux rope rose up and drove the standard flare reconnection. 

From t1 to t2, the HXR flux declined, the spectrum softened. We suggest the overlying dome-like structure constrained the rise of the flux rope, thereby suppressed the standard flare reconnection in the later phase of stage 1. During this period, the interaction between flux rope and the overlying dome-shaped structure triggered the QSL reconnection in fan-spine topology, leading to the brightening of ribbons S1 and R3--R6. The sequential brightening of these ribbons is displayed by the yellow patches of Figure~\ref{figure08}(b). At $\sim$t2, the AIA 1600~\AA{} emissions from ribbons R3--R6 reached their maximum values. The expanding overlying arcades eventually became invisible in EUV images, indicating the destruction of the fan-spine topology.

In stage 2,  The flux rope continued to rise after breaking the overlying magnetic confinement, facilitating its eruption (Figure~\ref{figure08}(c)) and caused a CME observed by LASCO C2. The standard flare reconnection continued and led to the re-enhancement of HXR emission after t2, with a secondary brightening of ribbons R1 and R2. At the second peak (t3), the HXR sources were also located at ribbons R1 and R2, and the spectrum slightly hardened compared to that at t2, suggesting a secondary energy release process.

%------**********************************--------------
\section{Summary and discussion} \label{Sect.5}

In this paper, we investigate an eruptive flare exhibiting quasi-parallel and semi-circular ribbons on 2023 September 20 using multi-wavelength data. The HXR light curve shows two major peaks temporally consistent with the sequential brightening of two quasi-parallel ribbons. At the local minimum between the two peaks, semi-circular ribbons brightened, followed by the expansion of large-scale arcades. We also reconstruct 3D magnetic field and find a dome-like structure associated with a fan-spine topology. We propose two reconnection processes:(1) Standard flare reconnection causes the two-stage brightening of quasi-parallel ribbons associated with two HXR peaks. (2) QSL reconnection in the fan-spine topology produces the semi-circular ribbon brightening. The interaction between the eruptive structure and the dome constrains the standard flare reconnection, leading to a distinct two-stage evolution.

\begin{figure}[htbp]
	\centering
	\includegraphics[width=0.95\textwidth]{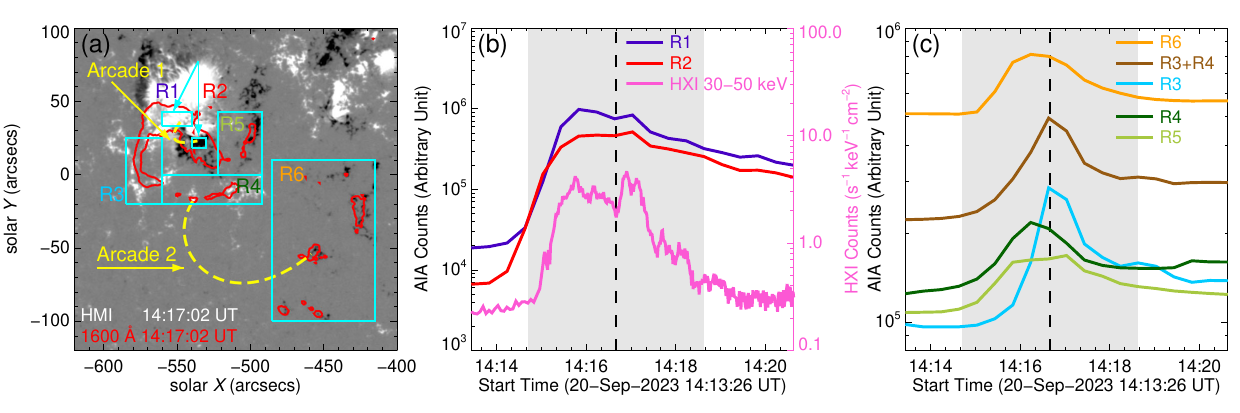}
	\caption{(a) HMI line-of-sight magnetogram overlaid with red contours at the 90\% intensity level of AIA 1600 {\AA} image at 14:17:02 UT. Cyan rectangles mark ribbon regions R1--R6. Arcade 1 and 2 are delineated by yellow dashed lines. (b) AIA 1600 {\AA} light curves in regions R1 (purple) and R2 (red), and 30--50 keV HXR flux (pink) from 14:13:26 to 14:20:38 UT. The gray shading represents the two-stage time interval, identical to that in Figure~\ref{figure01}. (c) Same as (b), but for regions R3 (cyan), R4 (olive), R3+R4 (brown), R5 (green), and R6 (orange). }
	\label{figure04}
\end{figure}

\begin{figure}[htbp]
	\centering
	\includegraphics[width=0.95\textwidth]{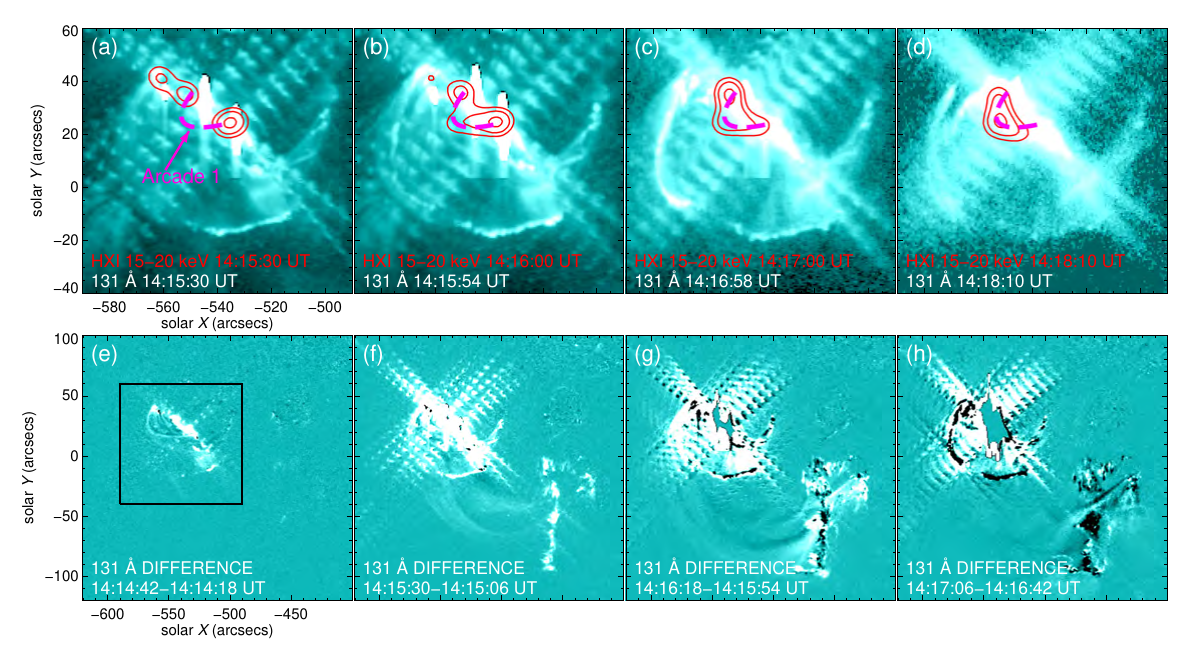}
	\caption{(a)--(d) Temporal evolution of AIA 131 {\AA} images and HXR 15--20 keV sources (red contours). ``Arcade 1'' denotes the post-flare loops as outlined by magenta dashed lines. (e)--(h) AIA 131 {\AA} running-difference images from 14:14:42 to 14:17:06 UT. The black box marks the field of view in panel (a). An animation of this figure is available.}
	\label{figure05}
\end{figure}

\begin{figure}[htbp]
	\centering
	\includegraphics[width=0.95\textwidth]{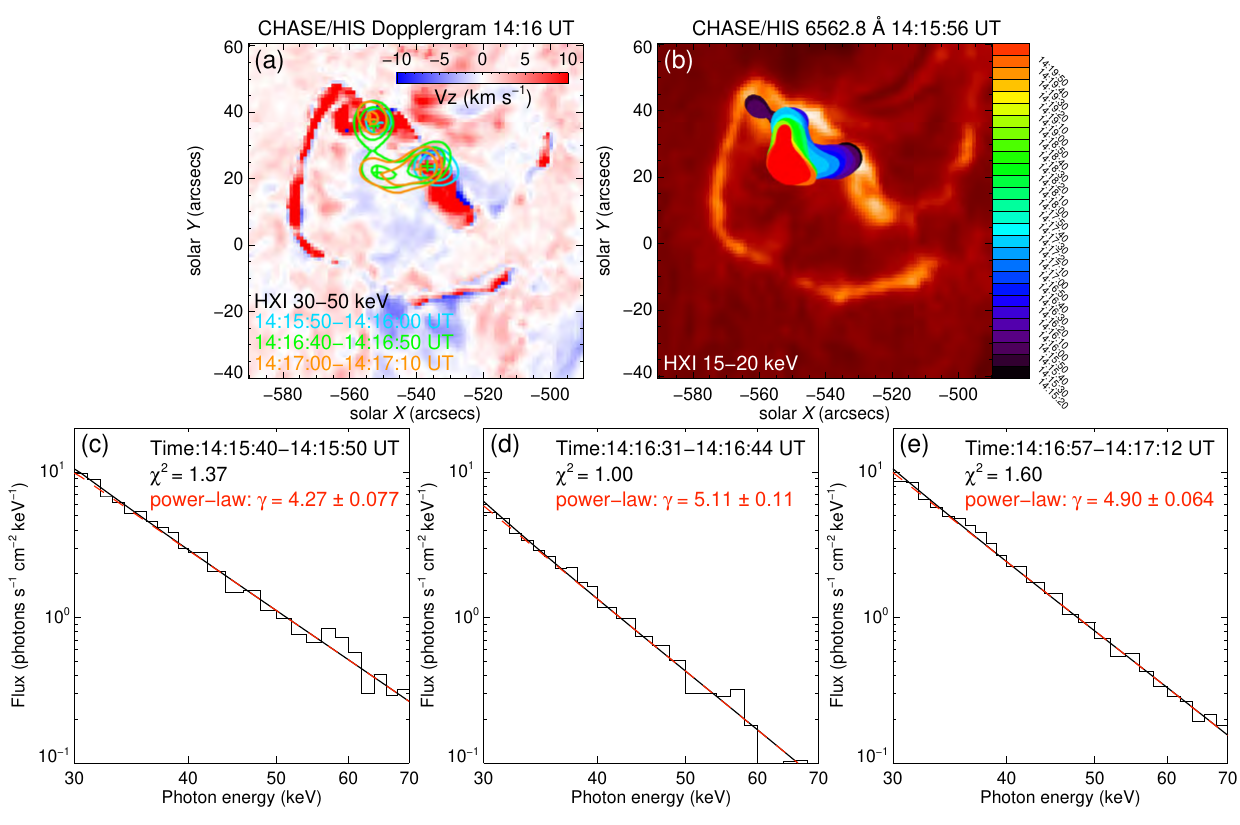}
	\caption{(a) HXR sources of 30--50 keV at 14:15:50--14:16:00 UT (cyan), 14:16:40--14:16:50 UT (green) and 14:17:00--14:17:10 UT (orange), superimposed onto the CHASE/HIS dopplergram at 14:16 UT.  The levels of contours are given by 40\%, 60\%, and 90\% of their respective maxima. The plus signs mark the local maxima. (b) CHASE H$\alpha$ image at 14:15:56 UT overlaid with the X-ray sources of 15--20 keV from 14:15:20 to 14:19:50 UT. Each source was reconstructed using a 10 s integration time. The contour level is given by 50\% of the maximum flux, filled with different colors corresponding to different time intervals. An animation of panel (b) is available. (c) HXI spectral fitting results during the interval of the first peak of HXR light curve. The integration time is from 14:15:40 to 14:15:50 UT. (d) and (e) Same as panel (c) but for the trough and the second peak of HXR light curve. The integration time is 14:16:35 to 14:16:45 UT and 14:16:57 to 14:17:07 UT, respectively. } 
	\label{figure06}
\end{figure}

\begin{figure}[htbp]
	\centering
	\includegraphics[width=0.95\textwidth]{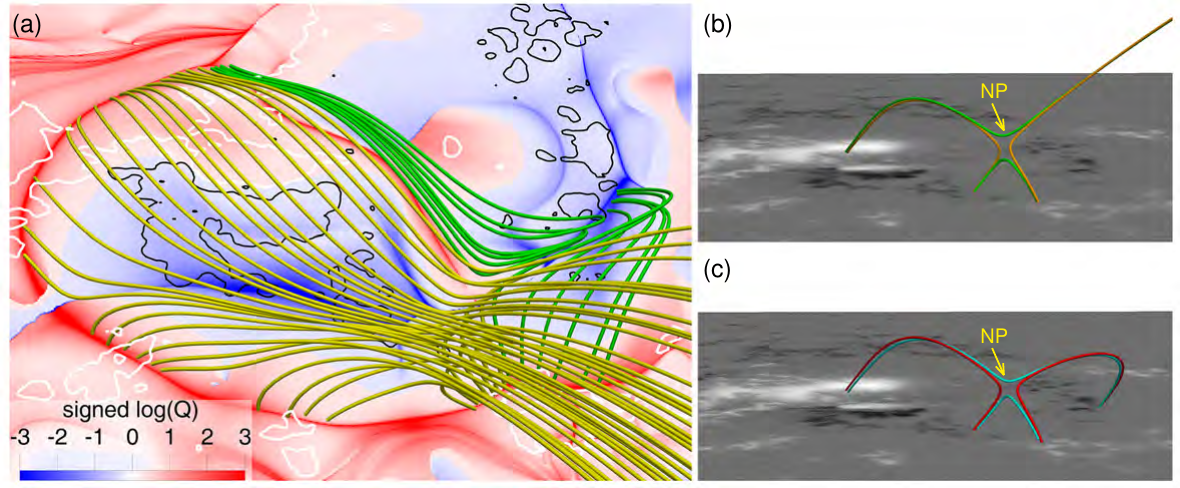}
	\caption{Selected field lines from the potential field extrapolation. (a) Yellow field lines depict a dome-like configuration and green field lines show another connectivity. The bottom boundary maps the distribution of the signed log(Q). The white and black contours show the photospheric vertical magnetic field of $\pm$ 300 G. (b) and (c) Two groups of null-point reconnection configuration. Orange lines in panel (b) and red lines in panel (c) represent field lines before the null point reconnection, while green lines in panel (b) and cyan lines in panel (c) represent field lines after it. Null point is marked as ``NP''.} 
	\label{figure07}
\end{figure}

\begin{figure}[htbp]
	\centering
	\includegraphics[width=0.98\textwidth]{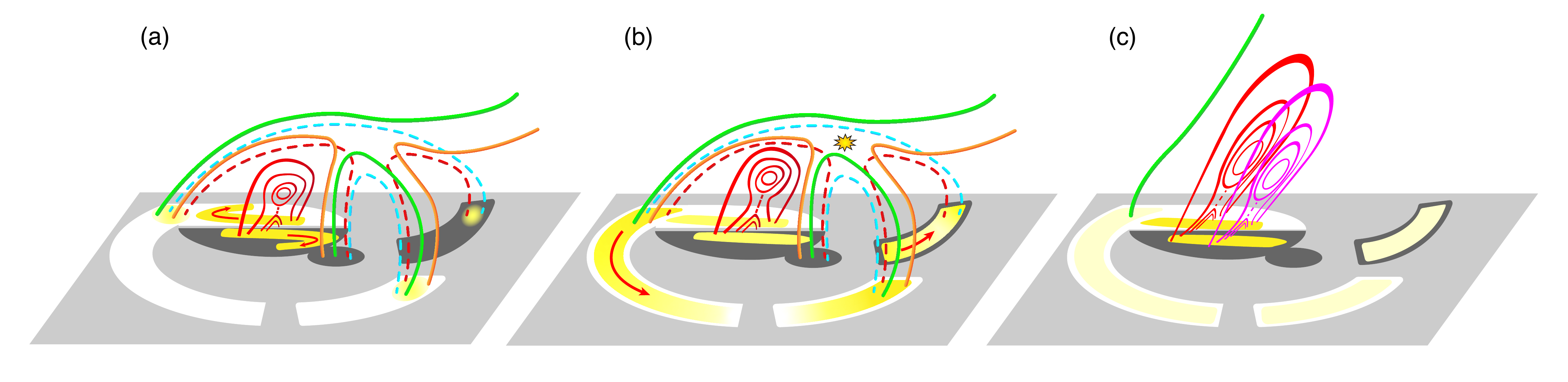}
	\caption{Schematic maps of the flare magnetic configuration around t1 (14:15:44 UT), t2 (14:16:40 UT), and t3 (14:16:53 UT) in panels (a)--(c), respectively. The white (black) area in the bottom represents the positive (negative) polarity. The fading yellow patches and red arrows show the the sequential brightening of flare ribbons. Green and orange lines have the same meanings as those in Figure~\ref{figure07}(b). Cyan and red field lines in Figure~\ref{figure07}(c) are changed into dashed lines here. The null point is marked with an asterisk in panel (b). Red and pink solid lines depict the cross section of the rising flux rope.} 
	\label{figure08}
\end{figure}

Previous studies have reported flares with both parallel and circular ribbons \citep{2017Masson,2019Liu,2020Liu,Joshi2021}, most of which are observed with the presence of a flux rope under dome-shaped structure. In eruptive events, the flux ropes rise and interact with the overlying dome structure, causing the null point reconnection. Such processes reduce the confinement on flux ropes and facilitate the eruption, known as the magnetic breakout model \citep{Sun2013,2016Chen}. A few of such flares are accompanied by failed eruptions, in which the flux rope is destroyed by reconnection with the dome structure \citep[see, e.g.,][]{2017Masson, Liu2018, Chen2021}. The flare we studied exhibits distinct two-stage features in the HXR emissions and ribbon dynamics. In particular, the peak flux of circular ribbons (R3+R4, the brown curve in Figure~\ref{figure04}(c)) temporally coincides with the flux minimum of quasi-parallel ribbons between the two HXR peaks. These features have not been reported in previous studies. We consider a combined effect of the two reconnection processes plays an important role in its eruptive nature.

According to the HXI light curve in 30--50 keV, there exist two minor peaks immediately behind the two major peaks. Although high-resolution imaging evidence is lacking, it could be interpreted as intermittent magnetic reconnection \citep[e.g.,][]{Wu2024}. \cite{2022Ning} reported an X-ray flux oscillation with a dominant period of about 20 seconds in the circular ribbon flare. They speculated that the flare quasi-periodic pulsations originate from repeated magnetic reconnection. 

According to the HXI imaging results, the HXR emissions are associated solely with the standard flare reconnection. No significant HXR emissions could be observed from the semi-circular ribbon or other locations. Note that it is a common phenomenon in the previous studies \citep{Sun2013,2015yangkai,2017Hernandez}. This may attribute to the limited dynamic range of the instrument. In this event, the HXR flux decreases to a local minimum at t2, suggesting that electron acceleration during fan-spine reconnection might be less efficient than that in the classical two-ribbon flare model. However, this interpretation requires  further validation through analysis of additional circular ribbon events with HXR emission and dedicated numerical simulations.

This event provides an unprecedented perspective to study the energy release of different types of magnetic reconnection in the same flare. However, we cannot distinguish the quantitative differences in their energy release processes. The flare was observed at the center of disk, while it is hard to image the faint coronal HXR sources with HXI observations, and the null-point configuration cannot be resolve. A stereoscopic observation from ASO-S and Solar Orbiter would be expected for further understandings of the 3D magnetic topology, making it possible to observe the footpoint and coronal structures in the same event.

\acknowledgments
We thank the CHASE, ASO-S/HXI, SUTRI, \textit{SDO}/AIA and HMI consortia for supplying the data.
The authors are supported by the National Natural Science Foundation of China (12203031, 12303061, 12333009), the National Key R\&D Program of China under grant 2022YFF0503002(2022YFF0503000), the Shandong Natural Science Foundation of China (ZR2023QA074), China Postdoctoral Science Foundation (2022TQ0189, 2023T160385, 2022M711931), and the Fundamental Research Funds for the Central Universities (KG202506).
The CHASE mission is supported by CNSA.
The ASO-S mission is supported by the Strategic Priority Research Program on Space Science of CAS, Grant No. XDA15320000. 
\textit{SDO} is a mission of NASA's Living With a Star Program.
SUTRI is a collaborative project conducted by the National Astronomical Observatories of CAS, Peking University, Tongji University, Xi'an Institute of Optics and Precision Mechanics of CAS and the Innovation Academy for Microsatellites of CAS.

\bibliography{reference}
\bibliographystyle{aasjournal}

%`````````````````````````````````````````````````````````
%`````````````````````````````````````````````````````````

%`````````````````````````````````````````````````````````
%`````````````````````````````````````````````````````````

%`````````````````````````````````````````````````````````
%`````````````````````````````````````````````````````````

%`````````````````````````````````````````````````````````
%`````````````````````````````````````````````````````````

%`````````````````````````````````````````````````````````

%`````````````````````````````````````````````````````````

%`````````````````````````````````````````````````````````
%`````````````````````````````````````````````````````````

%`````````````````````````````````````````````````````````
%`````````````````````````````````````````````````````````

\end{document}